\documentclass[fleqn,10pt]{wlscirep}
\usepackage[utf8]{inputenc}
\usepackage[T1]{fontenc}
\usepackage{bm}

\usepackage{amsmath}

\usepackage[normalem]{ulem}
\usepackage{braket}
\usepackage{color}
\usepackage{amssymb}
\usepackage{csquotes}
\usepackage[section]{placeins}\usepackage[justification=justified,singlelinecheck=false]{caption}

\DeclareUnicodeCharacter{2124}{$\mathbb{Z}$}

\usepackage[most]{tcolorbox}
\tcbuselibrary{breakable}
\newtcolorbox[auto counter]{boxX}[2][]{%
skin=enhanced jigsaw,breakable,colback=black!4,colframe=black!40,title=Sidebar \thetcbcounter: #2,#1}

\usepackage[backend=biber,style = nature]{biblatex}
\usepackage{orcidlink}

\title{Quantum Simulating Nature’s Fundamental Fields}

\author[1]{Christian~W.~Bauer\thanks{\tt cwbauer@lbl.gov}}
\author[2,3,4]{Zohreh Davoudi\thanks{\tt davoudi@umd.edu}\,\orcidlink{0000-0002-7288-2810}}
\author[5]{Natalie~Klco\thanks{\tt natalie.klco@duke.edu}\,\orcidlink{0000-0003-2534-876X}}
\author[6]{Martin~J.~Savage\thanks{\tt mjs5@uw.edu}\,\orcidlink{0000-0001-6502-7106}}
\affil[1]{Physics Division, Lawrence Berkeley National Laboratory, Berkeley, CA 94720, USA}
\affil[2]{Maryland Center for Fundamental Physics and Department of Physics, University of Maryland, College Park, MD 20742, USA}
\affil[3]{The Joint Center for Quantum Information and Computer Science (QuICS), NIST and University of Maryland, College Park, MD 20742, USA}
\affil[4]{The NSF Institute for Robust Quantum Simulation, University of Maryland, College Park, MD 20742, USA}
\affil[5]{Duke Quantum Center and Department of Physics, Duke University, Durham, NC 27708, USA}
\affil[6]{InQubator for Quantum Simulation (IQuS), Department of Physics, University of Washington, Seattle, WA 98195, USA}

\begin{abstract}
Simulating key static and dynamic properties of matter---from creation in the Big Bang to evolution into sub-atomic and astrophysical environments---arising from the underlying fundamental quantum fields of the Standard Model and their effective descriptions, lies beyond the capabilities of classical computation alone.
Advances in quantum technologies have improved 
control over quantum entanglement and coherence to the point where robust 
simulations are anticipated to be possible in the foreseeable future.
We discuss the emerging area of quantum simulations of Standard-Model physics, 
challenges that lie ahead, and opportunities for progress in the context of nuclear and high-energy physics.
\\
\\
\\
{\bf\large February 2023}
\end{abstract}

\begin{document}
\flushbottom
\maketitle
\thispagestyle{empty}

\section{The Quest for Simulating Nature's Quantum Fields}
\label{sec:intro} 
\begin{figure}[ht]
\begin{centering}
	\includegraphics[scale=0.92]{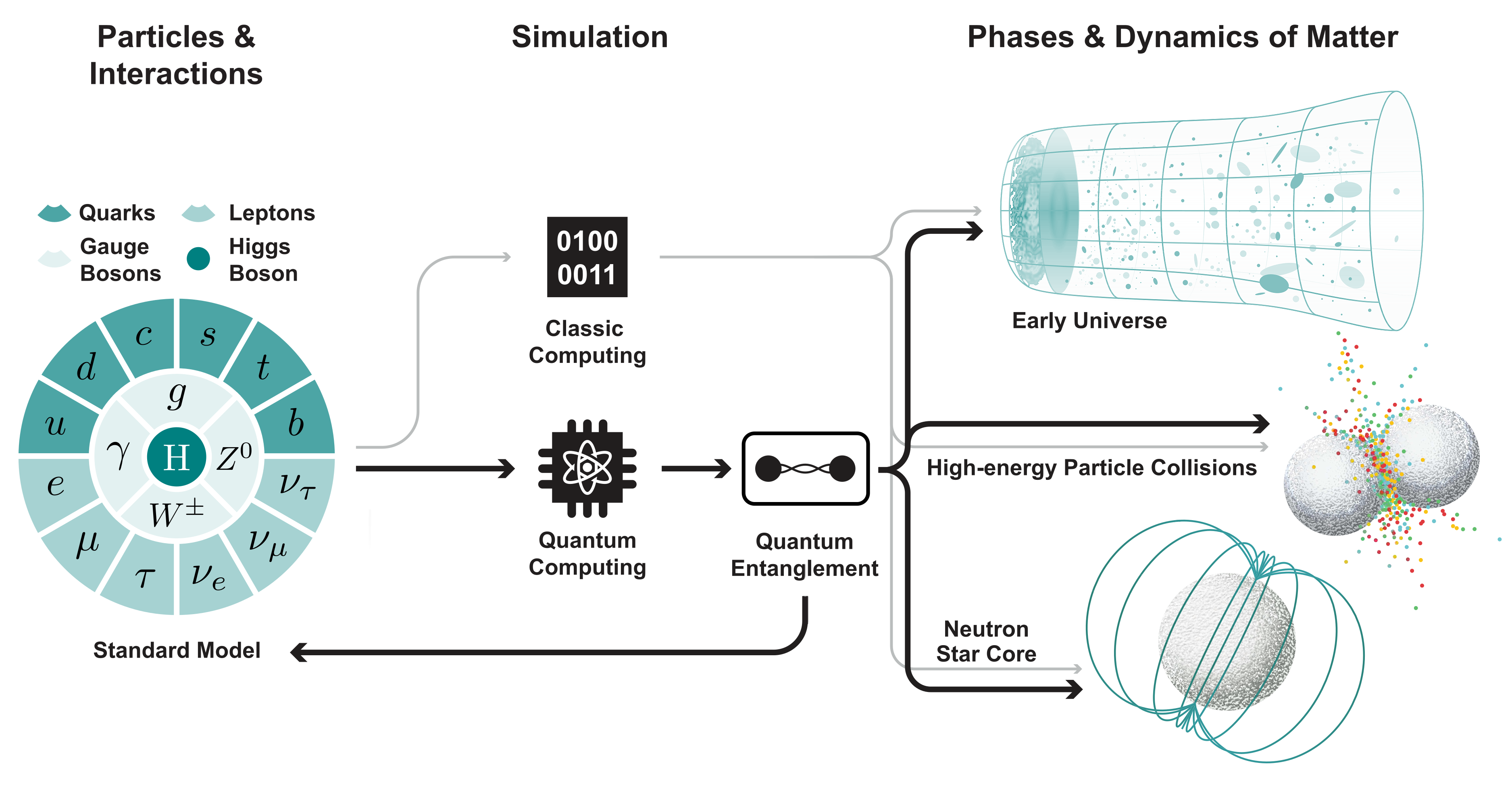}
	\caption{
	Simulating 
	the dynamics of extreme physical environments (right) emerging from the SM of quarks, leptons, gauge fields, and the Higgs boson (left) requires large-scale classical or quantum simulations (center). Quantum entanglement and coherence utilized by quantum computers are expected to enable progress while providing new insights into the SM itself. 
	}
	\label{fig:SMplus}
\end{centering}
\end{figure}

\noindent
The strong, weak, and electromagnetic forces of Nature are described by a theoretical framework called quantum field theory (QFT): a synthesis of quantum mechanics and relativity in which the finite speed of information is united with the quantum mechanical features of wavefunctions.
As the sub-atomic world became increasingly accessible in the laboratory, 
the discovery of new composite and elementary particles and their interactions
uncovered underlying global symmetries and local Abelian and non-Abelian gauge symmetries that now define the Standard Model (SM) of particle physics~\cite{Glashow:1961tr,Higgs:1964pj,Weinberg:1967tq,Salam:1968rm,Politzer:1973fx,Gross:1973id}.
The SM underlies a range of complex emergent phenomena in nuclear and high-energy physics, such as nuclei, particle-rich sprays in colliders or cosmic-ray atmospheric impacts,  nuclear reactions in terrestrial reactors or supernovae explosions, exotic phases of matter within dense celestial objects or details of gravitational waves from neutron-star mergers, 
and the formation of matter after the Big Bang.
Unfortunately,  simulations of many aspects of these systems and processes lie beyond classical computing, but an improved understanding of entanglement and the emergence of quantum simulation may provide a path forward, as depicted in Fig.~\ref{fig:SMplus}.

The non-Abelian gauge theory describing the strong force, Quantum Chromodynamics (QCD), results from invariance under local $SU(3)_C$ transformations
among the three \enquote{color} charges carried by each of the six known quark flavors and eight massless gluons mediating interactions.
In a landscape of quantum fluctuations and non-trivial topological features, the self-interactions among the gluons lead to the confinement of quarks and gluons into composite hadrons, hence the absence of free color charges in nature. 
The unified electroweak interactions are described by an $SU(2)_L\otimes U(1)_Y$ gauge theory with Yukawa interactions coupling the Higgs field to quarks and leptons. 
The Higgs vacuum-expectation value spontaneously breaks $SU(2)_L\otimes U(1)_Y\rightarrow U(1)_Q$,  
producing mass for the quarks, charged leptons, and three  weak gauge fields, $W^\pm$ and $Z^0$, 
while leaving the photon massless.

Despite the ubiquitous success of the SM in describing fundamental processes, it is known to be incomplete. Matter dominates over antimatter in the universe, but the SM does not possess sufficient CP-violation to produce it. Neutrinos have mass, despite being massless in the SM. 
There is clear astrophysical evidence for dark matter, however its potential particle content and interactions remain to be identified. 
Addressing such questions requires discovery of exotic effects traced to new particles and forces, 
perhaps through more-energetic particle colliders or through low-energy, high-intensity experiments that probe the effects of new physics in precision measurements.
As a result, searches for more fundamental fields of nature continue.

While some direct predictions of the SM are accessible by perturbative calculations, non-perturbative simulations are required for most processes, 
which involve configuration-space, i.e., Hilbert space, sizes that easily exceed the number of atoms in the universe.
The quantum-mechanical probability of a given process is determined by Feynman's path integral, summing the amplitudes of all possible trajectories weighted by the complex exponential of their action, $e^{iS}$. 
When $i S$ is a negative real number, these weights allow for systematic importance sampling central to lattice-field-theory programs on classical computers---enabling first-principles simulations of e.g., mesons, their decays and scattering, the muon's anomalous magnetic moment, properties of nucleons and nuclei, and the phase diagram of dilute matter at high temperature, see Refs.~\cite{FlavourLatticeAveragingGroupFLAG:2021npn,Davoudi:2022bnl,USQCD:2022mmc,Davoudi:2020ngi} for recent reviews. 
However, when $i S$ becomes complex, as  in finite-density systems and in real-time simulations, the sampling techniques can fail due to large cancellations. 
While ideas are being pursued to tame many \enquote*{sign problems}~\cite{Nagata:2021ugx,Bazavov:2019lgz,Alexandru:2020wrj}, they
are believed to be NP-hard~\cite{Troyer:2004ge}.

Unfortunately, sign problems arise frequently in nuclear and high-energy physics, challenging the pursuit of answers to many forefront questions, as identified and elaborated in recent studies, 
e.g., Refs.~\cite{NSAC-QIS-2019-QuantumInformationScience,Bauer:2022hpo,Catterall:2022wjq,Humble:2022klb}.
For example, first-principles predictions for phases and phase transitions of matter, e.g., those probed at the Large Hadron Collider (LHC) at CERN and the Relativistic Heavy Ion Collider (RHIC) at Brookhaven National Laboratory, and those relevant to the interior cores of neutron stars and the evolution of supernovae, are currently out of reach. 
The non-equilibrium and thermalization dynamics of matter produced during heavy-ion collisions and in the early universe also remain unresolved.
In particle colliders, available calculational strategies 
at high energies tend to break down at the lower energies relevant to fragmentation and the subsequent cascades of hadrons. 
In astrophysical environments, neutrinos play a significant role in transporting energy during core-collapse supernovae.
Background matter and neutrino densities influence the evolution of flavor as the neutrinos radiate from the core.
Given the resulting mixed-state entanglement structure of the neutrino fields, accurate simulations of these coherent processes is challenging.
Experimental programs probing the fundamental nature of neutrinos include searches for CP-violation in the Deep Underground Neutrino Experiment (DUNE) experiment, 
and for the violation of lepton number in $0\nu\beta\beta$-decay experiments in nuclei.
Simulations of nuclear-physics ingredients for these experimental programs present a significant challenge for classical computation.
Last, but not least, theories involving inflationary epoch of the universe, dark-matter models involving composite \enquote*{dark hadrons}, CP-violating scenarios occurring out of equilibrium in the early universe, or strongly-interacting QFTs living on the boundary of a bulk containing quantum-gravity models, all require simulations of various quantum fields in- and out-of-equilibrium, which are often intractable.

Amidst such challenges arose a key perspective:
Information is physical, computation is the science of information processing, and thus the intimate connection between simulation of physical systems and natural laws themselves provides a unique pathway for further fundamental discovery.   
In the early 1980s, these relationships led Feynman and contemporaries~\cite{Manin1980,Benioff1980, Feynman1982,Feynman1986,doi:10.1063/1.881299} to envision the next computational revolution, emboldened by the physics revelation that computation is reversible~\cite{5391327,5392446,Fredkin1982}.  
The idea was to introduce precisely controlled quantum degrees of freedom (d.o.f.) into a computational framework, naturally supporting Nature's complex many-body entangled states.
Quantum computers utilize interference through entanglement and coherence,
to efficiently explore large Hilbert spaces and perform highly parallel operations.
Processing information stored in quantum bits constitutes the first proposal satisfying Feynman’s \enquote{rule of simulation} that computational resources should be proportional to a system’s spacetime volume rather than to its exponential~\cite{Feynman1982}.  

\begin{boxX}[label={side:1}]{Ecosystem of quantum  architectures}
Access to a multi-architectural quantum ecosystem will be a hallmark of the quantum-simulation era. These architectures include photonics networks, trapped ions, superconducting radio-frequency cavities, 
neutral atoms in optical lattices or tweezer arrays, superconducting circuits,
as depicted in the figure, 
as well as molecular arrays, defects in solid state crystals, semiconductor technologies with embedded atomic spins, quantum annealers, coherent Ising machines with Floquet dynamics, Bose-Einstein condensates, and optical cavities.  A recent review of quantum-simulation architectures can be found in Ref.~\cite{Altman:2019vbv}.  
By isolating and connecting a few states in each Hilbert space, $\mathcal{H}$, from one or more quantum architectures, a dynamical quantum many-body system can be designed and simulated. 
Unique features of these architectures may pave the way toward large-scale scientific applications---from the long-range interactions of trapped ions, to the high local Hilbert space dimensionality in molecules and cavities, to the strong Rydberg blockade in atomic arrays. For the foreseeable future, noise-mitigation and error-correction strategies will have system and application dependence. 
Independent simulations on multiple architectures will be essential for providing verified and validated results with fully-quantified uncertainties.
\begin{center}
	\includegraphics[width=0.95\textwidth]{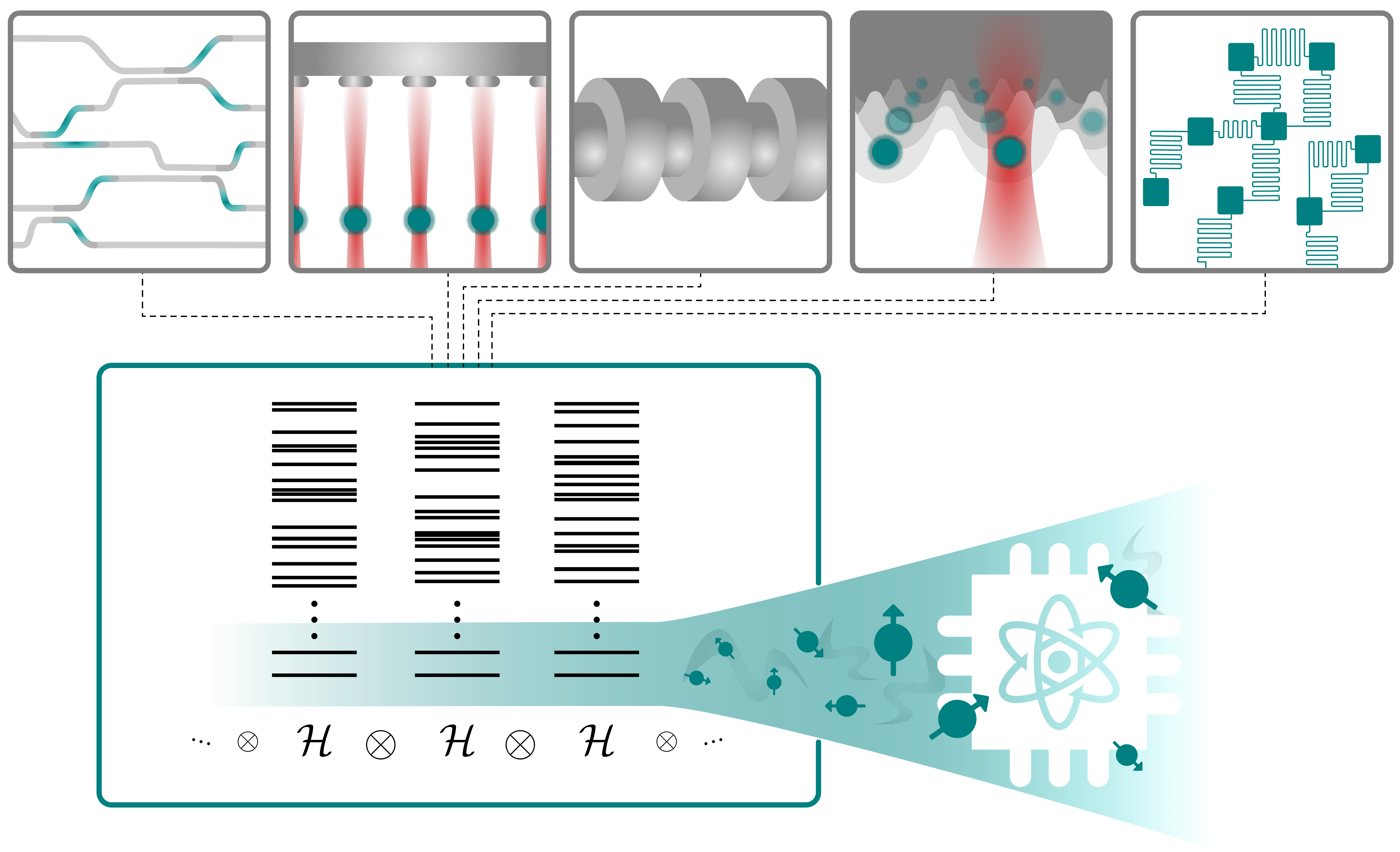}
\end{center}
\end{boxX}

While the vision of scientific quantum simulation continues to drive theoretical creativity, the current momentum of research is inspired by progress in controlling quantum few- and many-body systems. 
The vision expressed by Feynman is now becoming a reality.
With the identification of viable quantum architectures, see Sidebar~\ref{side:1}, valuable insights 
are being gained
into the natural quantum processing that 
are expected to underlie future technologies.  
For a review of quantum technologies in the simulation of gauge theories, see Ref.~\cite{Banuls:2019bmf}, and for additional discussions and reviews of advances since then, see e.g., Refs.~\cite{Kasper:2020akk,Aidelsburger:2021mia,Klco:2021lap,Bauer:2022hpo}. 
Quantum-simulating platforms can operate in various modes: analog, digital, or hybridizations of both.  
By experimentally designing devices governed by desired Hamiltonians, analog simulators implement time evolution naturally and continuously. 
It is perhaps this promising strategy that most closely aligns with Feynman’s vision of a computer acting \enquote{exactly the same as Nature}~\cite{Feynman1982}.
While Abelian interactions can be engineered in analog simulators~\cite{Mil:2019pbt,Yang:2020yer,Zhou:2021kdl,schweizer2019floquet, Gorg:2018xyc}, the intricacy of non-Abelian symmetries and running of the strong coupling raises the question whether 
the full range of SM Hamiltonians will be available in analog simulators given that some subatomic components, such as quarks and gluons, cannot be \enquote{externally controlled} due to confinement. 
In other words, 
can colorless objects be manipulated in the laboratory to mimic the dynamics of colorful quarks and gluons?
Despite recent progress, the answer to this important question is presently unknown---justifying the simultaneous exploration of digital quantum simulations, which rely  on the 
observation
that the rules of quantum mechanics are the same at all scales.
However, digital designs can challenge Feynman’s vision of predominantly local interactions, specifically avoiding a \enquote{very enormous computer with arbitrary interconnections throughout the entire thing}~\cite{Feynman1982}.  
For this reason, and others, promising proposals are emerging for simulating quantum fields through hybrid digital-analog simulation schemes that
leverage the flexibility of digital control, the native interactions of analog evolution, and the natural statistics of fundamentally fermionic/bosonic computational elements.

\section{From Quantum Fields to Quantum Simulators}
\noindent
Simulating quantum fields with finite computational resources, quantum or classical, often requires two layers of truncation: one to relinquish the continuity of infinite space by utilizing a finite-size spatial lattice, and one to digitize continuous fields and enforce a cut off at each spacetime point. 
The continuum limit of a lattice simulation is recovered by approaching a second-order phase transition where correlation lengths diverge, allowing physical volumes to be represented by increasing quantum complexity capable of supporting the many-body entanglement that is necessary for long-distance correlations~\cite{Klco:2021biu}.
As such, from capturing entanglement present in the continuum limit, to the delocalization of fermions with lattice chiral symmetry~\cite{Kaplan:1992bt, Kaplan:1992sg, Narayanan:1993zzh, Narayanan:1993ss, Shamir:1993zy}, these considerations provide hints that a form of non-locality may need to be incorporated into quantum simulations of fields, following the recent quantum-information advances in condensed-matter research, e.g. Refs.~\cite{augusiak2012many, zeng2015quantum,Savary2017Liquids, Sachdev:2018ddg}.
A classical computational method that demonstrates clear advantages from entanglement considerations is tensor networks~\cite{White:1992zz,Rommer:1997zz,Vidal2003Slightly,Vidal:2003lvx,Pichler:2015yqa,Kuhn:2015zqa,montangero2018introduction,Tilloy:2018gvo,silvi2019tensor,Banuls:2018jag,Banuls:2019rao,Banuls:2019qrq,Emonts:2020drm,Meurice:2020pxc,Meurice:2022xbk,Banuls:2022vxp}. 
In this framework, precision simulations of low-dimensional systems can be performed by providing efficient representation of a quantum state based on the entanglement entropy distributed throughout the system.
Remarkable progress has been made in tackling some complex problems in quantum many-body systems and field theories, and the range of problems that may be amenable to solution using tensor networks continues to  grow.
Nonetheless, dynamical processes of the fields can produce abundant spacelike entanglement, 
e.g., the quantum correlations generated by inelastic scattering processes, 
rendering tensor network simulations 
increasingly difficult with increasing energy and complexity~\cite{Milsted:2020jmf}.
In another example of the role of entanglement considerations in quantum simulations,
correlation structures can guide state preparation of massive fields, whose exponentially decaying two-point functions can be translated into corresponding hierarchies in quantum operations~\cite{Klco:2019yrb,Klco:2020aud}.
This allows classical simulations in small spatial volumes to inform the design of quantum circuits that can be used to prepare a state in volumes not accessible classically.
Extending ongoing entanglement-guided design of tensor networks and of quantum circuits~\cite{Klco:2019yrb,Klco:2020aud}, as our understanding of entanglement in nuclear and high-energy physics~\cite{Ho:2015rga,Kharzeev:2017qzs,Baker:2017wtt,Cervera-Lierta:2017tdt,Beane:2018oxh, Beane:2018oxh, Beane:2019loz,Tu:2019ouv,Beane:2020wjl,Beane:2021zvo,Kharzeev:2021yyf,Robin:2020aeh,Low:2021ufv,Gong:2021bcp,Roggero:2021asb,Johnson:2022mzk} and in QFT~\cite{Reeh1961,summers1985vacuum,summers1987bell1,summers1987bell2,VALENTINI1991321,Srednicki:1993im,Holzhey:1994we,Halvorson:1999pz,Audenaert:2002xfl,Reznik:2002fz,Reznik:2003mnx,Calabrese:2004eu,2004PhRvA..70e2329B,Retzker_2005,kofler2006entanglement,Ryu:2006bv,Marcovitch:2008sxc,Calabrese:2009ez,Calabrese:2009qy,Casini:2009sr,Zych:2010yk,Calabrese:2012nk,Calabrese:2012ew,Ghosh:2015iwa,Soni:2015yga,Dalmonte:2017bzm,Witten:2018zxz,Mendes-Santos:2019tmf,DiGiulio:2019cxv,Klco:2020rga,Kokail:2020opl,Roy:2020frd,Klco:2021cxq,Mueller:2021gxd,Klco:2021biu,Dalmonte:2022rlo}
improves, so too will our ability to simulate the SM.

A local Hamiltonian formulation of U(1) and SU(N) lattice gauge theories was developed by Kogut and Susskind (KS)~\cite{Kogut:1974ag, Banks:1975gq}. 
In this formulation,
fermions  are distributed over a staggered lattice such that each site hosts a one-component fermion, and gauge bosons are placed on the links between them, as illustrated in Sidebar~\ref{side:2}. In the electric basis, the link Hilbert spaces contain representations of the underlying gauge group, e.g., integer values labeling the one-dimensional representations for U(1), $2j+1$-dimensional spin-$j$ representations for SU(2), and higher dimensional tensor representations for SU(3).  
Transitions within a link Hilbert space from the actions of the link operator, $\hat{U}$, are indicated by arrows within each representation diagram in Sidebar~\ref{side:2}. 
The Hamiltonian  is composed of local gauge-invariant operators representing the energy stored in the electric fields (on the links), energy stored in the magnetic fields 
(plaquette operators forming a closed loop of gauge links), 
fermion kinetic energy and fermion-gauge-field interaction energy, and fermion mass terms. 
Local gauge invariance is enforced on the states by a set of local Gauss's law operators $\hat{G}^a(\mathbf{x})$ for the $a^{\rm th}$ generator of the group, which commute with the Hamiltonian. The physical Hilbert space includes only the states that satisfy
$\hat{G}^a(\mathbf{x})|\Phi\rangle_{\rm phys} = 0$. 
This implies that
the electric-field fluxes entering and exiting a site are balanced by the corresponding charge present at the site. 
The continuum and infinite-volume limits of simulations need to be taken, and the gauge-boson truncation to be extrapolated away, to obtain physical observables. 
Efforts to understand and realize such limits in the Hamiltonian lattice-field-theory frameworks are underway, as are efforts to port, rather than recreate, the  techniques developed
for classical lattice-field-theory calculations~\cite{Klco:2018zqz,Briceno:2020rar,Carena:2021ltu,Ciavarella:2021lel,Clemente:2022cka,Farrell:2022wyt}. 
Alternate formulations of quantum fields 
within and beyond the KS framework have been proposed~\cite{Byrnes:2005qx,Ciavarella:2021nmj,Klco:2019evd,Zohar:2014qma,Zohar:2019ygc,Zohar:2019ygc,Anishetty:2009nh,Raychowdhury:2019iki,Davoudi:2020yln,Kaplan:2018vnj,Haase:2020kaj,Bauer:2021gek,Ji:2020kjk,Brower:1997ha,Alexandru:2019nsa,Singh:2019uwd,Kreshchuk:2020dla,Buser:2020cvn,Ciavarella:2022zhe,Bauer:2021gek,Kane:2022ejm},
with a goal to reduce the simulation resources required to make predictions at a given level of precision.

\begin{boxX}[label={side:2}]{Foundations of the KS Hamiltonian and its mapping to quantum devices}
\begin{center}	\includegraphics[width=0.99\textwidth]{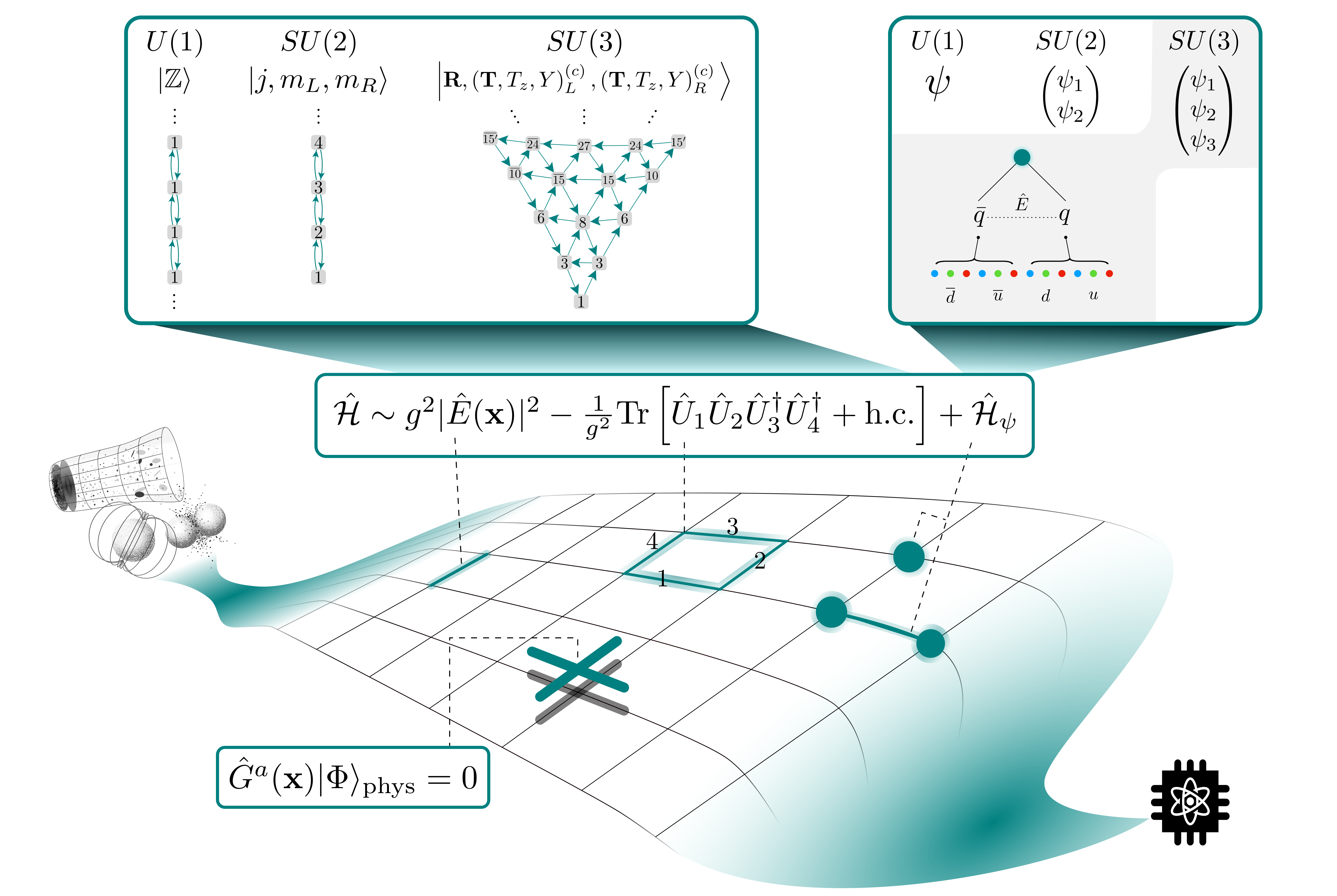}
\end{center}
In the KS Hamiltonian, lattice elements 
include the link operator $\hat{U}$,
plaquette operator~(\includegraphics[width = 5mm]{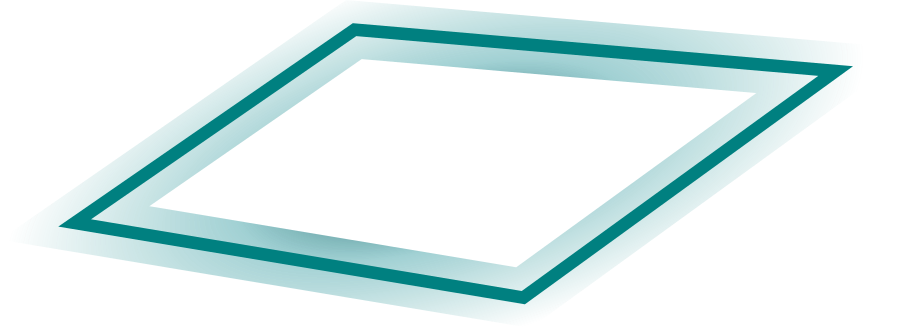}), fermion ($\hat{\psi}$, \includegraphics[width = 2.5mm]{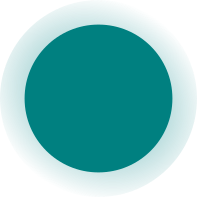}) kinetic, interaction~(\includegraphics[width = 5mm]{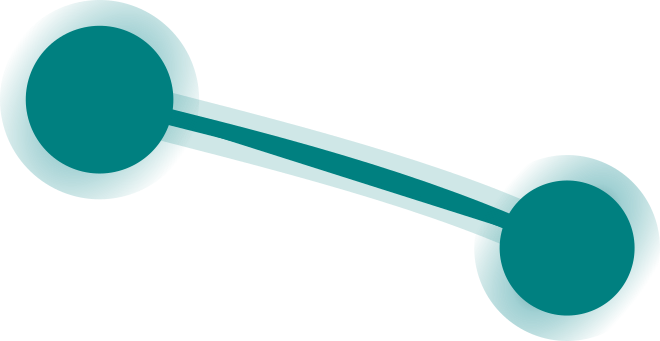}), and mass terms, and the Gauss's law constraint~(\includegraphics[height = 2.5mm]{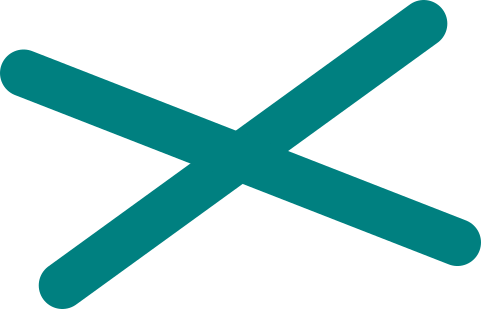}).
The connectivity and dimensionality of the electric ($E$) basis representations is utilized to digitize links, as proposed by Byrnes and Yamamoto~\cite{Byrnes:2005qx}.
For SU(3), link states that run between 
sites 
are
$|{\bf R}, (T, T^z, Y)_L^{(c)}, (T, T^z, Y)_R^{(c)} \rangle$, 
where the labels define orientations in the left and right color spaces, 
denoted by color-isospin, its third component, and color-hypercharge.
${\bf R}$ is the SU(3) irreducible representation of the link (common to both ends), 
which can also be denoted by the number of up and down indices
($p,q$) of the associated tensor representation.
The maximum values of $T_{L,R}$, $T_{L,R}^z$ and $Y_{L,R}$ are bounded for a given ${\bf R}$.
For a 3+1D simulation, each lattice site contains the ``intersection'' of
three left-states and three right-states that combine to satisfy Gauss's law.
The plaquette operators act on four link states around a closed path, transforming as a Gauss's law preserving color singlet at each lattice site, 
but acting as a ${\bf 3}$ or $\overline{\bf 3}$ on each link.
Each of the integers defining the state of each link are mapped to qubit registers.
For example, accommodating irreps with  $p,q \le 2$  requires
two registers with two qubits.  
The maximal isospin in these irreps is $T=2$ with $|T_z|\le 2$
and $|Y|\le 2$, each requiring its own register of qubits.
On one spatial (two 
staggered) sites for two-flavor (up and down quarks) QCD, the Hilbert space content of $\hat{\mathcal{H}}_\psi$ is mapped to 12 qubits~\cite{Farrell:2022wyt}.
\end{boxX}

The basic workflow of a quantum simulation consists of three parts: preparation of a non-trivial initial state, unitary time evolution, and measurement of desired physics observables. 
For example, when aiming to simulate particle collisions, 
this workflow may translate to the preparation of wavepackets on top of a non-trivial vacuum, evolving the wavepackets to a region where they interact and subsequently separate, and finally measuring  properties of asymptotic states, as would be observed by laboratory detectors~\cite{Jordan:2011ci}.
The preparation of arbitrary initial states is generically QMA-Complete~\cite{Kempe2003QMA3local,Kempe2004LocalHam,Oliveira2005latticeQMA}.
Recent progress in developing state preparation techniques include adiabatically approaching states of a target theory from those of a simpler theory~\cite{Calabrese:2006rx}, adiabaticity-violating processes such as quenched dynamics,  imaginary-time 
evolution~\cite{Motta2020Imaginary,McArdle2019Imaginary,YeterAydeniz2020Imaginary,Liu:2020eoa}, and variational procedures that may 
incorporate global entangling operators~\cite{Peruzzo2014Variational,McClean2016Theory,kandala2017hardware}.  Quantum algorithms for preparing thermal states or evaluating thermal expectation values for field theories have also been developed~\cite{deJong:2021wsd,Czajka:2021yll,Davoudi:2022uzo}. 
However, more research is needed to demonstrate the applicability of these methods for preparation of states in the SM, e.g., hadrons and nuclei that are composites of quarks and gluons. 
Measurement protocols allow for extraction of physical observables.  In general, access to the full final-state wavefunctions is not possible, but recently proposed techniques, such as classical shadows~\cite{Aaronson2018Shadow,Huang2020Predicting,Kokail:2020opl,Elben:2022jvo}, permits sampling of the density matrix and evaluation of observables with significant reductions in required resources.

When digital quantum computers are used,   the time-evolution operator is discretized commonly using a strategy called Trotterization. 
The Hamiltonian is broken into simpler terms, 
and the system is evolved separately under those terms for short time intervals to mitigate systematic errors from the non-commutativity among terms. 
The circuit design for each separate evolution, called a Trotter step, is more manageable and can benefit from classical pre-processing.  Currently,  circuit design in this context refers to decomposing operations into single- and two-qubit universal gate sets,
but it is expected that (hardware-specific)
multi-qubit abstractions will be important in future developments of simulations for gauge theories. Implementing many Trotter steps consecutively results in longer simulation times with an improvable accuracy, see e.g., Refs.~\cite{Byrnes:2005qx,Jordan:2011ci,Barata:2020jtq,Byrnes:2005qx,Shaw:2020udc,Kan:2021xfc,Ciavarella:2021nmj,Lamm:2019bik,Paulson:2020zjd,Liu:2020eoa,Kreshchuk:2020dla,Stryker:2021asy,Davoudi:2022xmb} for recent progress in digital algorithms for field theories. 
Gauss's law in gauge-theory simulations may be violated 
by approximate algorithms and experimental noise, but it can be enforced via various techniques~\cite{Martinez:2016yna,Stryker:2018efp,Raychowdhury:2018osk,Stannigel:2013zka,Klco:2019evd,Kasper:2020owz,Halimeh:2020ecg,Tran:2020azk,Lamm:2020jwv,Nguyen:2021hyk}.
Analog hardware platforms implement time evolution continuously, and while engineering the Hamiltonian of gauge theories is generally challenging, progress is being made, see e.g., Refs.~\cite{Zohar:2012xf,Banerjee:2012pg,Tagliacozzo:2012df,Stannigel:2013zka,Zohar:2013zla,Hauke:2013jga,Kuhn:2014rha,Kasper:2015cca,Zohar:2015hwa,Mezzacapo:2015bra,Bazavov:2015kka,Yang:2016hjn,Gonzalez-Cuadra:2017lvz,Gorg:2018xyc,Davoudi:2019bhy,Surace:2019dtp,Luo:2019vmi,schweizer2019floquet,Mil:2019pbt,Yang:2020yer,Ott:2020ycj,Kasper:2020akk,Dasgupta:2020itb,Andrade:2021pil,Aidelsburger:2021mia,Osborne:2022jxq}.
When Gauss's law is simple, efficient mappings can be made to quantum simulators, e.g., the nearest-neighbor blockade interactions among Rydberg atoms can be used to suppress violations of Gauss's law in quantum link models (QLM) of U(1) gauge theories ~\cite{Surace:2019dtp}. 
There are proposals for implementations of non-Abelian Gauss's law, e.g., mapping to the internal levels of cold atoms engineered such that Gauss's law correspond to conservation of angular momentum in atomic collisions~\cite{Zohar:2013zla}. 
However, the generation of gauge-matter or plaquette operators is challenging because three- and higher-body interactions are less natural than two-body interactions in analog simulators. 
This has led to the integration of digital simulation strategies in analog simulations. While it may not be possible to engineer the dynamics under the full Hamiltonian in a single application, the evolution can be decomposed into separate terms, each implemented efficiently using intrinsic d.o.f. and interactions in the simulator. Examples are: leveraging auxiliary d.o.f. in layers of optical lattices that couple to the main d.o.f. via two-body interactions only~\cite{Zohar:2016wmo,Zohar:2016iic,Bender:2018rdp}, 
or the use of tunable coupled trapped-ion-phonon interactions to simulate fermion-boson field dynamics~\cite{Davoudi:2021ney,Zhang:2016lyo}, 
or by separately simulating electric and magnetic Hamiltonians with the natural multi-body interactions of Rydberg atoms~\cite{Gonzalez-Cuadra:2022hxt}.  While encoding the gauge-field d.o.f. in the target theory to bosonic d.o.f. in the simulator could, in principle, eliminate the need for truncating the gauge-field Hilbert space, in practice, the dynamics may need to be limited to a finite (but still sizable) number of bosonic excitations. 

On present-day quantum hardware, error mitigation is essential~\cite{Bennett:1996gf,Dankert2009Exact,Dur2005Standard,Emerson2007Symmetrized,Temme2017Error,li2017efficient,Endo2018Practical,Kandala2019Error,He:2020udd,viola1999dynamical,souza2012robust,suter2016colloquium,ARahman:2022tkr,Urbanek:2021oej,Zhang:2021lzr,Leyton-Ortega:2022xfu,Martinez:2016yna,Klco:2019evd,Nguyen:2021hyk,Tran:2020azk,Nguyen:2021hyk}, and error correction is not yet possible due to decoherence introduced by the deep circuits required. 
Progress is being made toward identifying robust error-correction schemes that take advantage of structure and gauge-field redundancies inherent in the gauge-theory encoding to decrease the error-correction qubit overhead~\cite{Klco:2021jxl,Rajput:2021trn}. 
Noise-mitigation and error-correction schemes tailored to gauge-theory simulations will continue to be developed and benchmarked in the coming years. 

\section{Emerging Quantum Simulations of the Standard Model Fields}
\noindent 
\noindent 
In the current period of developments of quantum simulations for the SM, 
many studies are focused on identifying and estimating the quantum resource requirements of problems that are classically intractable 
but are anticipated to benefit from quantum advantages~\cite{NSAC-QIS-2019-QuantumInformationScience,Bauer:2022hpo,Catterall:2022wjq,Humble:2022klb}. 
In parallel, small simulations have been run on available quantum hardware in an effort to understand the capabilities of present devices, and guide
hardware and algorithmic developments. 
While the focus of this article is on simulating the fundamental fields of the SM,
simulations of its descendant Effective Field Theories (EFTs) and phenomenological models utilize similar techniques to advance high-energy and nuclear physics.
For low-energy nuclear physics, much of the complexity that drives systems beyond classical computing 
emerges at the level of nucleons (and mesons).
As such, there is a growing effort to simulate nuclear EFTs,  
shell models, effective model spaces for nuclei, and more, see e.g., Refs.~\cite{Roggero:2019myu,Holland:2019zju,Roggero:2020sgd,Stetcu:2021cbj,Choi:2020pdg,Baroni:2021xtl,Turro:2021vbk}.
As with the fundamental fields, understanding and utilizing 
the entanglement structure of nuclei and systems of nucleons is key to success,
and early progress in these directions is encouraging, see e.g., Refs.~\cite{Johnson:2022mzk,Robin:2020aeh,Faba:2021kop,Kruppa:2021yqs}.
The rate of progress and the rapidly increasing number of studies precludes a comprehensive review of the literature here, and progress in just a few representative areas will be summarized.

\begin{figure}[t!]
\centering
\includegraphics[width = 0.99\textwidth]{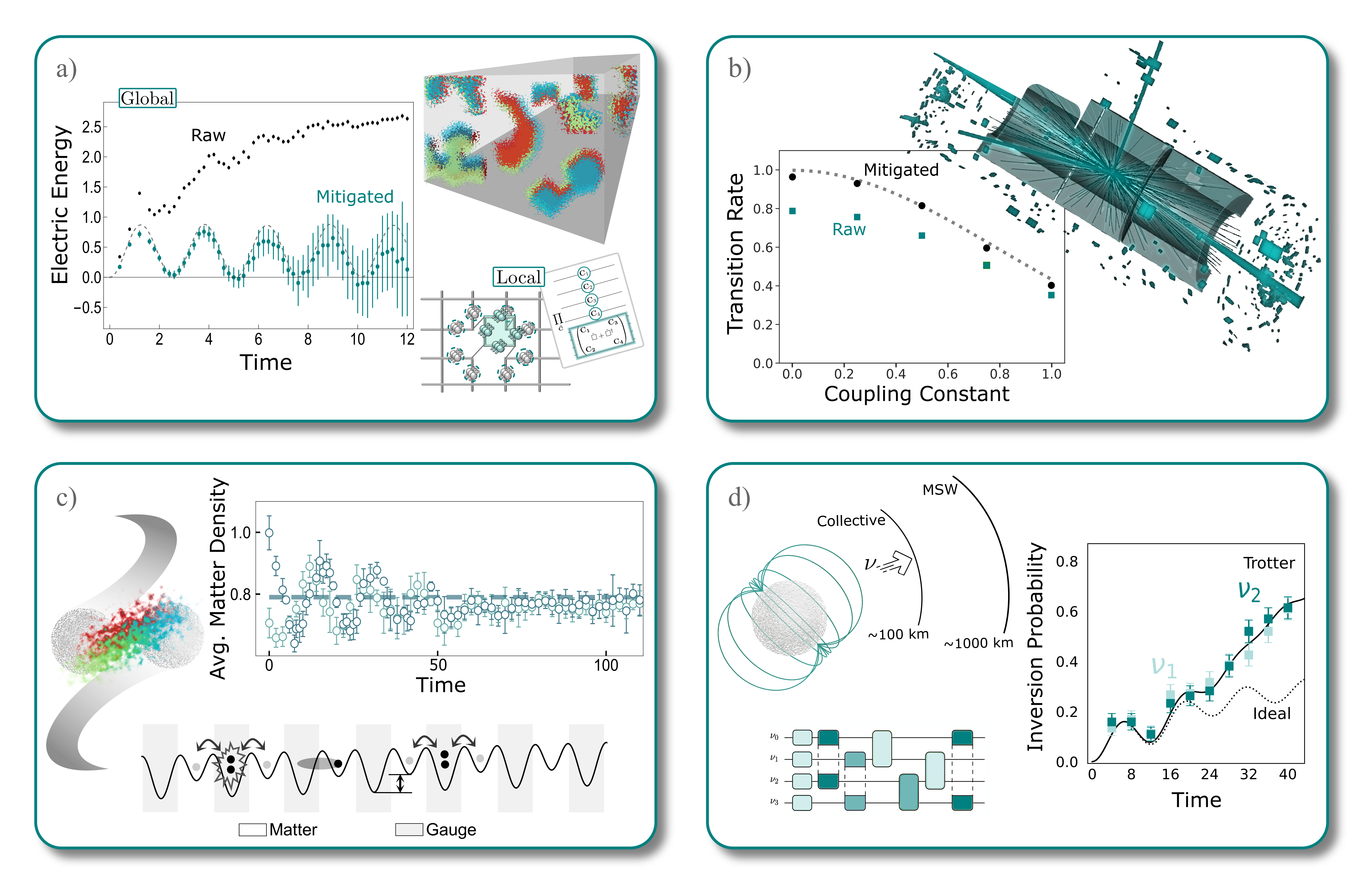}
	\caption{
	a) The electric energy of a single plaquette of 
    SU(3) Yang-Mills gauge theory 
    (in a global basis) 
    under time-evolution of up to 100 Trotter-steps,
    including the ${\bf 1}, {\bf 3}, \overline{\bf 3}$ and ${\bf 8}$ irreducible representations of SU(3), calculated on  {\tt ibmq\_Manila},
    and associated scalable local-basis circuit construction~\cite{Ciavarella:2021nmj}.
	b) Vacuum-vacuum transition rates induced with Wilson-line operators, relevant for evaluating the soft functions, is obtained for a 1+1D scalar field theory for three lattice sites and two qubits per site using {\tt ibmq\_Manhattan}~\cite{Bauer:2021gup}.
	c) The local matter density after a quantum quench of two initial states (depicted with different colors) in a 1+1D QLM is evaluated on a 71-site optical-lattice quantum simulator~\cite{Zhou:2021kdl}.
	The quantity asymptotes to a constant value, and the corresponding temperatures match.
	d) The inversion probability of two of the four neutrinos coherently evolved is evaluated as a function of Trotterized time using Quantinuum {\tt H1-1} trapped-ion computer, and the associated circuit~\cite{Amitrano:2022yyn}.
	}
	\label{fig:Qsimplus}
\end{figure}
%

\vspace{0.2 cm}
\noindent
\textbf{Lattice-gauge-theory simulations.}
The first efforts toward lattice-gauge-theory simulations using quantum hardware began in 2016 with a pioneering paper by Martinez {\it et al.}~\cite{Martinez:2016yna} 
that simulated time evolution in 1+1D quantum electrodynamics (the Schwinger model) for a small spatial lattice. 
The Schwinger model continues to be a "sandbox" for ideas and experiments~\cite{Martinez:2016yna,Klco:2018kyo,Kokail:2018eiw,Nguyen:2021hyk} for quantum simulation 
as it shares a number of  features analogous to those of QCD. 
Quantum simulations of non-Abelian lattice gauge theories have also  been performed on small lattices, 
such as SU(2) and SU(3) theories in 1+1D. 
For example, the first steps toward preparing vacuum and single-hadron states using the 
Variantional Quantum Eigensolver (VQE) algorithm have been taken~\cite{Atas:2021ext,Farrell:2022wyt,Farrell:2022vyh}. 
A small number of quantum simulations of lattice gauge theories in 2+1D have been performed, but simulations in 3+1D remain to be done. 
One of the major differences compared with 1+1D simulations is that the gauge fields are dynamical and 
are only partially constrained by Gauss's law, and their encoding 
requires significant resources.
Furthermore, few-plaquette systems in SU(2) and SU(3) theories have been simulated~\cite{Klco:2019evd,ARahman:2021ktn,ARahman:2022tkr,Ciavarella:2021nmj,Ciavarella:2021lel}.
Current focus is on developing efficient encodings for initializing and evolving states in Yang-Mills gauge theory and, in particular, implementation of Gauss's law. 
By integrating over the gauge space at each lattice site, the qubit requirements of multi-plaquette simulations can be reduced significantly, see the example depicted in Fig.~\ref{fig:Qsimplus}. 
Efficient state-preparation strategies in non-Abelian gauge theories in 2+1D are also being  actively investigated~\cite{Ciavarella:2021lel}.

\vspace{0.2 cm}
\noindent
\textbf{Fragmentation in particle colliders.} 
High-energy colliders are one of the most direct tools to test our understanding of fundamental interactions at short distances. 
Many scattering processes at high energies can be factorized into short-distance contributions that are perturbative and can be computed using analytical and numerical tools, 
and long-distance contributions that are non-perturbative and are often classically inaccessible. 
Therefore, quantum computers may be used to solve only for the long-distance contributions. 
Separating long- and short-distance dynamics from one another is achieved using EFTs. For collider physics, a relevant EFT is Soft-Collinear Effective Theory (SCET)~\cite{Bauer:2000yr}, and the main non-perturbative ingredients required for most observables are given by operators formed out of energetic fields moving in the same direction (collinear), 
or soft functions composed out of fields whose energy is small (soft). 
A quantum algorithm for 
the soft function has been worked out in detail for a scalar field theory~\cite{Bauer:2021gup}, 
which shares many of the features 
relevant for gauge theories, see Fig.~\ref{fig:Qsimplus}.  See also Refs.~\cite{Bepari:2020xqi,Bepari:2021kwv} for other algorithms developed for this application. 
These proof-of-principle simulations open the door to further theory and algorithmic studies for the case of QCD soft functions relevant at the LHC, and for other non-perturbative collider-physics quantities, 
such as structure functions (e.g., parton distribution functions)~\cite{Lamm:2019uyc,Echevarria:2020wct,Li:2021kcs,Mueller:2019qqj,Perez-Salinas:2020nem,Qian:2021jxp,Pedernales_2014}.

\vspace{0.2 cm}
\noindent
\textbf{Thermalization dynamics.} Heavy-ion colliders 
aim to re-create the conditions after the Big Bang to shed light on phases of matter at extreme densities and temperatures. 
Among the key open questions are~\cite{Berges:2020fwq,Lovato:2022vgq} how  the initial conditions of collisions
influence the approach to thermalization, 
what sets the time scale of thermalization, 
what the signatures of non-equilibrium phases of matter are, 
and how  they are imprinted in the final-states after hadronization. 
Ideally, the experiments would be replicated 
on a quantum simulator
by colliding energetic composite hadrons and nuclei.
Such complex state preparation is beyond current reach,
but simpler quenched-dynamics simulations can be performed today 
to create non-equilibrium conditions in experiments.
As an example, the far-from-equilibrium dynamics of a QLM description of the 
Schwinger model
were studied in Ref.~\cite{Zhou:2021kdl} using a 
Bose-Hubbard quantum simulator. 
As depicted in Fig.~\ref{fig:Qsimplus}, it was shown that thermal equilibrium emerges at late times after a quantum quench in this model.
Another scenario that could drive the system toward thermal equilibrium is coupling to environment, or coupling to extra d.o.f. during  simulations, as investigated in Ref.~\cite{deJong:2021wsd}. 
The role of entanglement in equilibration and thermalization~\cite{kaufman2016quantum,geraedts2016many} of gauge theories~\cite{Mueller:2021gxd}, and a better understanding of  phenomena such as the slow or potential lack of thermalization of quantum many-body \enquote{scar} states~\cite{turner2018weak,serbyn2021quantum},
and dynamical quantum phase transitions in non-equilibrium settings~\cite{heyl2018dynamical, heyl2013dynamical, zhang2017observation, guo2019observation}, and their prevalence in gauge theories~\cite{Aramthottil:2022jvs,Desaules:2022kse,Halimeh:2022rwu,Zache:2018cqq, Mueller:2022xbg,VanDamme:2022loc, VanDamme:2022rnb, Jensen:2022hyu}, have  begun to generate excitement in the field.

\vspace{0.2 cm}
\noindent
\textbf{Collective neutrino oscillations.} 
In core-collapse supernovae, the density of neutrinos is large enough that  their flavor evolution is modified by coherent interactions among themselves, and thus changing  the transport of energy, momentum, and flavor during the explosion. 
The relevant energies and densities are much lower than the scale of weak interactions, 
and the flavor evolution can be captured using an effective theory~\cite{Pantaleone:1992xh,Pantaleone:1992eq} 
in which $N$ neutrinos with two flavors
are described by $N$ spin-${\frac{1}{2}}$ d.o.f. (and three flavors maps to spin-$1$ d.o.f.)
interacting  
via a Heisenberg Hamiltonian with coefficients encoding neutrino density and momentum spectrum.
The vacuum flavor oscillations are captured by including an external magnetic field. 
Significant advances in our understanding of neutrino dynamics in such extreme environments 
have uncovered surprising complexity, 
including a dynamical phase transition separating fast and slow modes~\cite{Friedland:2003eh,Bell:2003mg,Sawyer:2004ai,Rrapaj:2019pxz,Cervia:2019res,Martin:2021bri,Roggero:2021fyo,Roggero:2022hpy}, 
and an appreciation for the role of entanglement.
It is now recognized that quantum simulations are essential in developing robust predictive 
capabilities for these systems~\cite{Hall:2021rbv,Yeter-Aydeniz:2021olz,Illa:2022jqb,Amitrano:2022yyn,Illa:2022zgu}, see Fig.~\ref{fig:Qsimplus} for an example. 
A more comprehensive understanding of entanglement structure in these systems may emerge~\cite{Cervia:2019res,Roggero:2021asb,Amitrano:2022yyn,Illa:2022zgu,Roggero:2022hpy} 
as quantum computers become more capable in the coming years.

\section{Moving ahead: Challenges, Opportunities, and Perspectives}
\noindent
The challenges arising in the pursuit of reliable quantum simulations of SM physics are substantial.
While establishing a quantum advantage for any scientific application will 
be a historic feat, such an achievement is only an early step along a long path necessary to realize the envisioned potential of quantum simulations.
Fortunately, opportunities are numerous not only upon arrival of large-scale reliable quantum simulators/computers, but also throughout the journey of development. These include a modern focus on the role of entanglement and its resource interpretation, 
exploring qualitatively new observables in experiment, 
devising low- and high-precision calculations of quantities inaccessible to classical computing, 
guiding the design of classical algorithms to incorporate fundamental non-localities,
the development of new EFTs extending the predictive capabilities of quantum simulations,
impacting the design of quantum-simulating hardware, improving the functionality and applicability of quantum sensors, 
and beyond.

There is consensus in some aspects of the trajectory to large-scale quantum simulation, such as integrating both classical and quantum computers in future computing ecosystems.
However, the frequency of emerging new ideas suggests that the ultimate path may yet to be formulated.
Rather than narrowing the considered landscape, recent modest simulations and resource estimates of gauge theories have served to expand and illuminate the practicality of this endeavor, supporting the importance of theoretical flexibility in field representations and time-evolution protocols to mirror the growing diversity of quantum hardware.
Furthermore, though it may prove to be a significant consideration in deciding the path forward, initializing scattering states for QCD---composites of quarks and gluons forming hadronic wavepackets---has not yet been achieved except in highly truncated low-dimensional cases.
Lastly, for the initial design of simulation strategies and subsequent  analyses, experience with classical calculations of lattice QCD indicates that combining simulations with designer EFTs can  accelerate connections to experiment. Examples are mitigating discretization  uncertainties 
or using chiral perturbation theory to extend accessible parameter ranges, see e.g., Ref.~\cite{Bernard:2002yk} for a discussion.
In fact, such frameworks have the potential to allow for efficient quantum simulations to be connected to observables that are formally inefficient for quantum computers, introducing another important layer of consideration beyond complexity classes.

Simulations of chiral gauge theories---in which the distinct chiralities of fermions interact differently with the gauge fields, as is the case for the electroweak interactions in Nature---has so far eluded efficient classical simulation. 
One hope is that  efficient quantum algorithms for chiral gauge theories can be found, so as to allow, for example, simulations of weak-scale baryogenesis with controlled uncertainties.
There is much to be gained by simulating models that share features with theories of interest like QCD, both to develop methods for computing dynamical quantities, and for realizing the quasi-local connectivity required for the 3+1D field-theory simulations.
Importantly, such simulations may provide qualitative understanding of unexplored dynamics of strongly interacting matter that could point to new avenues for experimental and theoretical investigation.

The algorithms and techniques developed to simulate the SM fields concern a diverse set of interactions, both fundamental and effective, and span orders of magnitude in energy.  
As in the case of classical computing~\cite{Christ2011},
the co-design of quantum simulations, including the quantum hardware tailored for nuclear physics and high-energy physics applications, e.g., multi-qubit gate operators required for magnetic operations in non-Abelian gauge theories, 
will  find application in other domains sciences, 
such as materials science, quantum chemistry, error-correction, and other areas of quantum information sciences.  
The algorithms and techniques that are central to  quantum simulation may also be utilized to encode (and decode) quantum sensors into distributed entangled states for enhanced sensitivity~\cite{Davis2016Approaching,Zhou:2017kxr,Ahmed:2018oog,Zhuang:2019pzt,Guo2020Distributed,Kaubruegger2021Variational,Marciniak:2021tld,Xia2021Enhanced,Hernandez-Gomez:2021osr,Alderete:2022zrf,Brady:2022bus}. 
The entanglement structure of the SM fields and their interactions may be particularly valuable for determining response function of quantum sensors to external fields.

For the quantum simulation of SM physics to be successful, robust quantum error correction is expected to be required.
As such, error-corrected digital quantum computers with universal instruction sets are considered to be the ultimate technology target for robust simulation. 
Nonetheless, the once-distinct development paths of analog and digital technologies are converging in some systems, providing opportunities for advantageous hybridization, e.g., incorporating analog features for representing Fermi and Bose statistics in a digital computer or digitally implementing strong many-body interactions beyond those \enquote{nearby} or accessible  \enquote{designer} Hamiltonians in available architectures.
Further afield in speculation are opportunities to utilize key aspects of the SM, e.g., topological structures and fundamental symmetries, as error-correction protocols themselves.

Future quantum simulations represent our most promising route to understanding the dynamics of complex many-body systems that lie far beyond the capabilities of classical computing, including the in- and out-of-equilibrium dynamics of dense matter.
They are anticipated to enable predictions of new phenomena, to improve understanding of observations, and to point the way forward for new experimental programs.  
The increasing focus in quantum correlations and entanglement, the key to the additional capabilities of quantum computers, is already providing new insights
and understandings of the complex systems emerging from the SM.
We expect that the next decades will see remarkable breakthroughs and advances in nuclear physics and high-energy physics resulting from the embrace of entanglement and the pursuit of quantum simulation, building on Feynman's vision.

\section*{Acknowledgements}

We would like to thank all of our collaborators and colleagues 
for shaping our perspectives of the evolving field of quantum simulation for nuclear physics
and high-energy physics research.
Christian Bauer acknowledges  support by the Director, Office of Science, Office of
High Energy Physics of the U.S. Department of Energy under the
Contract No. DE-AC02-05CH11231, in particular through Quantum Information Science Enabled Discovery
(QuantISED) for High Energy Physics (KA2401032).
Zohreh Davoudi acknowledges support by the U.S. DOE’s Office of Science Early Career Award, under award no. DE-SC0020271, National Science Foundation Quantum Leap
Challenge Institute for Robust Quantum Simulation (\url{https://rqs.umd.edu}) under grant OMA-2120757, Maryland Center for Fundamental Physics (\url{https://mcfp.physics.umd.edu}) at the University of Maryland, and the DOE’s Office of Science, Office of Advanced Scientific Computing Research, Quantum Computing Application Teams program, under fieldwork proposal number ERKJ347, and the Accelerated Research in Quantum Computing program under award DESC0020312.
Martin Savage acknowledges  support by U.S. Department of Energy, 
Office of Science, Office of Nuclear Physics, InQubator for Quantum Simulation (IQuS)\footnote{\url{https://iqus.uw.edu/}} under Award Number DOE (NP) Award DE-SC0020970 via the program on Quantum Horizons: QIS Research and Innovation for Nuclear Science\footnote{\url{https://science.osti.gov/np/Research/Quantum-Information-Science}}, 
the Department of Physics 
(\url{https://phys.washington.edu}) and the College of Arts and Sciences 
(\url{https://www.artsci.washington.edu})
at the University of Washington.

\section*{Image Credits}
\textbf{Figure 1} resulted from a collaboration between Zohreh Davoudi and Chad Smith, Data Visualization and Multimedia Designer at University of Maryland, College Park.
\textbf{Image in Sidebar 1} resulted from a collaboration between Natalie Klco and Samantha Trieu, Graphic Designer at the Lawrence Berkeley National Laboratory.
\textbf{Image in Sidebar 2} resulted from a collaboration between Natalie Klco and Samantha Trieu.  
Icons at the far left and right were contributed by Chad Smith. 
The top-left box reproduces content from Ref.~\cite{Ciavarella:2021nmj} with permission from Anthony Ciavarella.  
The top-right box reproduces content from Ref.~\cite{Farrell:2022wyt} with permission from Roland Farrell.
\textbf{Figure 2} is a compilation of results and graphics from Refs.~\cite{Amitrano:2022yyn,Bauer:2021gup,Zhou:2021kdl,Ciavarella:2021nmj,Ciavarella:2022zhe}. 
Specifically, the results in the \emph{top-left panel} are a version of a figure from Ref.~\cite{Ciavarella:2021nmj} updated with current technologies and error mitigation strategies, and the circuit diagram is from Ref.~\cite{Ciavarella:2022zhe}, both reproduced with permission from Anthony Ciavarella.
The event display in the \emph{top-right panel} was generated by the ATLAS Collaboration (\url{https://atlas.cern/updates/press-statement/shedding-new-light-higgs}) run number: 209381, event number: 72873013, date: 2012-08-21 04:17:16 CEST. 
The plot in the same panel is reproduced from Ref.~\cite{Bauer:2021gup} 
with permission from Christian Bauer.   
The results and the optical superlattice-simulator diagram in the \emph{bottom-left panel} are reproduced from Ref.~\cite{Zhou:2021kdl} with permission from Zhao-Yu Zhou. 
The content of the \emph{bottom-right panel} is reproduced from Ref.~\cite{Amitrano:2022yyn} with permission from Valentina Amitrano, and the neutron-star graphic is contributed by Chad Smith.

\printbibliography

\end{document}